# WAKEFIELD BAND PARTITIONING IN LINAC STRUCTURES

R.M. Jones[†], V. Dolgashev[†], K.L.F. Bane[†], and E. Lin[‡]
Stanford Linear Accelerator Center, P.O. Box 20450, Stanford University, CA 94309

## Abstract

In the NLC project multiple bunches of electrons and positrons will be accelerated initially to a centre of mass of 500 GeV and later to 1 TeV or more. In the process of accelerating 192 bunches within a pulse train, wakefields are excited which kick the trailing bunches off axis and can cause luminosity dilution and BBU (Beam Break Up). Several structures to damp the wakefield have been designed and tested at SLAC and KEK and these have been found to successfully damp the wakefield [1]. However, these $2\pi/3$ structures suffered from electrical breakdown and this has prompted us to explore lower group velocity structures operating at higher fundamental mode phase advances. The wakefield partitioning amongst the bands has been found to change markedly with increased phase advance. Here we report on general trends in the kick factor and associated wakefield band partitioning in dipole bands as a function of phase advance of the synchronous mode in linacs. These results are applicable to both TW (travelling wave) and SW (standing wave) structures

Paper presented XXI International Linear Accelerator Conference (LINAC2002),
Hotel Hyundai in Gyeongju, Korea.
August 19 – August 23rd, 2002

[†] Supported by Department of Energy grant number DE-AC03-76SF00515
[‡] Supported by Office of Science Energy Research Undergraduate Laboratory Fellowship by the Department of Energy, visiting from the University of Texas at Austin

# WAKEFIELD BAND PARTITIONING IN LINAC STRUCTURES

R.M. Jones[†], V. Dolgashev[†], K.L.F. Bane[†], and E. Lin[‡]; SLAC, ARD-A, Menlo Park, USA

Abstract

In the NLC project multiple bunches of electrons and positrons will be accelerated initially to a centre of mass of 500 GeV and later to 1 TeV or more. In the process of accelerating 192 bunches within a pulse train, wakefields are excited which kick the trailing bunches off axis and can cause luminosity dilution and BBU (Beam Break Up). Several structures to damp the wakefield have been designed and tested at SLAC and KEK and these have been found to successfully damp the wakefield [1]. However, these $2\pi/3$ structures suffered from electrical breakdown and this has prompted us to explore lower group velocity structures operating at higher fundamental mode phase advances. The wakefield partitioning amongst the bands has been found to change markedly with increased phase advance. Here we report on general trends in the kick factor and associated wakefield band partitioning in dipole bands as a function of phase advance of the synchronous mode in linacs. These results are applicable to both TW (travelling wave) and SW (standing wave) structures

## 1. INTRODUCTION

The initial NLC/JLC structures, built and tested at SLAC and KEK, were designed to accelerate electron and positron beams up to 0.5 TeV and up to 1.5 TeV in later upgrades. In order to efficiently accelerate the charged particles, multiple bunches are accelerated within a pulse of the RF field and this requires loaded field gradients of the order of 70MV/m. These accelerating structures were chosen to be 1.8 meters in length as this limits the number of fundamental and higher order mode couplers needed per structure. The dipole mode frequencies of these structures were detuned in order to ensure destructive interference of dipole wakefield modes excited by the beam. The wakefield that is seen by bunches trailing the exciting bunches is known as the long-range wakefield. The wakefield in the structure DDS1 (Damped Detuned Structure) was well-predicted by a circuit model [2] and subsequently measured in ASSET [3]. Further structures were fabricated and tested in which both the efficiency of transference of energy to the electron beam and the wakefield damping were optimized. However, a beam-based experiment revealed a substantial and unpredicted phase shift of the accelerating mode along the structure. This prompted a full autopsy of the structures and it revealed breakdown had been occurring and this had been concentrated in the high group velocity end of the structures. This has led to a series of new, shorter, low group velocity test structures (the "T" series [4]) which are constant gradient and are not provided with wakefield damping. Recent high gradient tests performed on these structures were very encouraging as the breakdown rates were found to be substantially reduced. We have two further series of structures which will incorporate damping of the dipole modes: SW accelerators which operate in the $\pi$ mode and travelling wave accelerators which operate at a $5\pi/6$ phase advance per cell.

The new SW structure consists of 15 cells and will eventually incorporate both detuning of the dipole frequencies and damping of the wake by incorporating a limited number of choke mode cavities in the structure [5]. The travelling wave structures are similar in design to the original DDS series except that they are shorter by a factor of up to 3 and they accelerate the beam with a reduced group velocity (3% compared to 12% in DDS). This series of accelerator structures is know as the "H series" because they operate at a higher phase advance ($5\pi/6$ compared to $2\pi/3$) which is needed to reduce the group velocity whilst at the same time preserved the same average iris radius. We maintain the average iris radius (radius/accelerating mode wavelength = $a/\lambda \sim 0.18$) in order to keep the wakefield along the bunch (the intra-bunch or short range wake) to acceptable levels. For the short NLC bunches the average strength of the slope of the short-range wakefield is proportional to the -3.8 power of the iris radius.

A comparison of the dipole wakefield band structure for a TW $2\pi/3$ phase advance structure with a $\pi$ SW detuned structure is presented in the following section and the damping requirements that this requires are discussed. In section 3 the general properties of band portioning of kick factors are presented.

## 2. UNCOUPLED ANALYSIS OF TRANSVERSE WAKEFIELDS

The transverse wakefield excited by a particle beam can be decomposed into modes which kick the beam transversely to the axis of acceleration. Here, we use an uncoupled analysis in which we calculate the wakefield at the synchronous frequency using the individual cell kick factors. This analysis is valid to a good approximation for the first few meters behind the driving bunch. At longer distances a coupled mode [6] or spectral function [7] analysis must be used. For an N-cell accelerating structure the envelope of the wakefield at a distance s behind the first bunch we derive as the absolute value of a summation:

[†] Supported by Department of Energy grant number DE-AC03-76SF00515

[‡] Supported by Office of Science Energy Research Undergraduate Laboratory Fellowship by the Department of Energy, visiting from the University of Texas at Austin

$$W(s) = 2\left|\sum_{n=1}^{N} K_n \text{Exp}\left[j\frac{\omega_n s}{c}(1+\frac{j}{2Q_n})\right]\right| \quad (2.1)$$

where for the $n^{th}$ mode, $K_n$ is the transverse kick factor, $\omega_n/2\pi$ is the synchronous frequency and $Q_n$ is the quality factor of the mode. A modal expansion similar to eq. (2.1) is also found in [8]. The kick factor is evaluated as:

$$K_n = \frac{|\int_L E_z \text{Exp}[j\omega_n s/c]dz|^2}{4\frac{\omega_n a_n^2}{c}U_n L(1-\frac{v_{gn}}{c})} \quad (2.2)$$

Here, $a_n$ is the radius of the $n^{th}$ iris, L the periodic length of the cell, Ez is the on-axis electric field and $U_n$ is the energy stored per cell in a mode. The kick factor also depends on the group velocity $v_{gn}$ [9] and provided the synchronous phase is close to $\pi$ then the group velocity dependence will be a negligible correction and it can be ignored. For all DDS structures is has indeed been found to be a small correction. However, for the SW and new high phase advance structures this is no longer the case. We calculated the kick factors and wakefields for a representative TW detuned structure known as DS1 and for a 8 SW structures each of which consists of 15 cells and they are both detuned with a 10% bandwidth. The 8 different SW structures effectively make up a 120 cells structure. We require 8 structures as the detuning provided by 15 cells in one structure alone is insufficient.

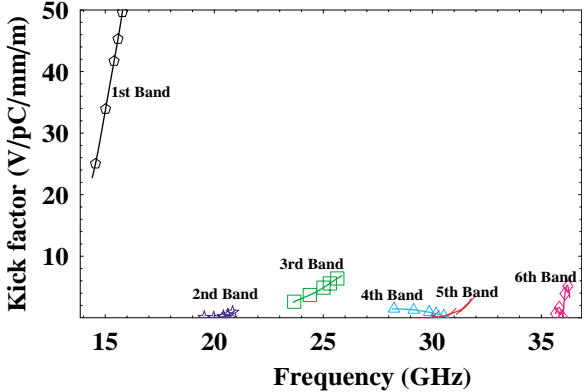

Figure 1. Kick factors in a travelling wave accelerator: DS1 (Detuned Structure). The complete set of 206 kick factors is obtained by interpolation from the calculation of kick factors for 5 cells (shown with dots). The largest kick factors are all concentrated in the first band. The third and sixth bands, although they are almost an order of magnitude smaller than the first, also affect the beam dynamics in the linac. All of these three bands must be detuned.

The results of these calculations, performed with HFSS and GdfidL [10], are shown in Figs 1 and 2 respectively. For the standing wave structure the kick factors are no longer linearly dependent on the synchronous frequency and they are not concentrated in only the first band. The wakefield that results from each of these bands for the travelling wave detuned structure and the standing wave structure is shown in Fig 3 and 4 respectively. In this calculation we have not included the effects of the finite group velocity of the dipole mode.

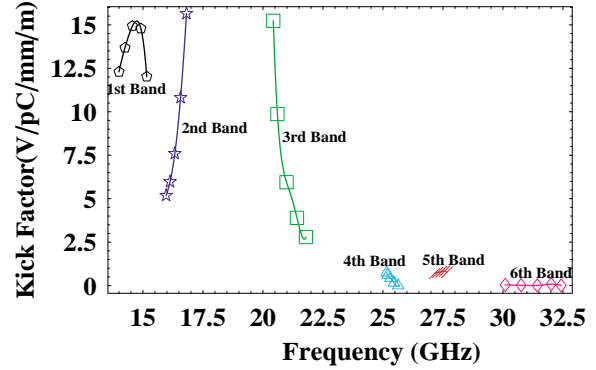

Figure 2. Kick factors in standing wave accelerator design SW1. The complete set of 120 kick factors is obtained by interpolation from the calculation of kick factors for 5 cells (shown with dots). The first three bands kicks are of similar order of magnitude and they all must be damped and detuned. The $4^{th}$ and $5^{th}$ bands are an order of magnitude smaller than the first three but they also must considered in a full analysis of the beam dynamics as they also contribute to BBU instability.

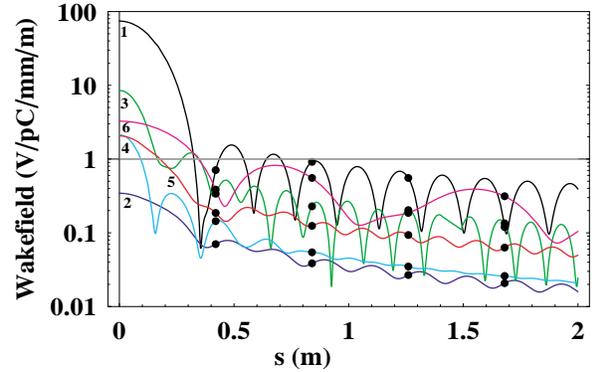

Figure 3. Individual bands, ranging from the first to the sixth of the envelope of the wakefield corresponding to the kick factors of the travelling wave structure given in Fig 1. The dots are positioned at the location of each individual bunch (spaced by 42cm). Four out of a total 191 trailing bunches are shown

Beam dynamics studies [11] indicate that the wakefield must be below unity in order that the BBU instability not be an issue. The wake at the position of the bunches is shown in Fig 3 for the travelling wave structure and it is clear that the wake remains below unity at these locations and thus BBU is unlikely to be a problem. The $3^{rd}$ and $6^{th}$ bands have significant kick factors compared to the first band and these modes were detuned by enforcing and

Erf variation to the iris thickness of all cells. The wakefield for the SW structure shown in Fig 4 reveals that the first three band are all equally important and consequently they all must be carefully damped. The wake at the position of the first trailing bunch is below unity for all bands apart from the third band. The third band requires additional detuning in order to accelerate the rate of decay of the wake.

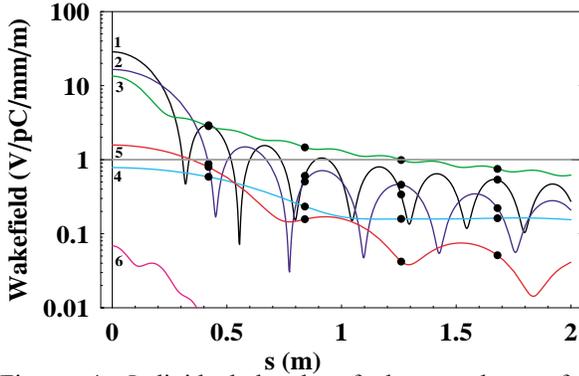

Figure 4. Individual bands of the envelope of the wakefield corresponding to the kick factors of the standing wave structure given in Fig 2.

## 3. GENERAL PROPERTIES REGARDING BAND PARTITIONING OF KICK FACTORS

In order to assess the behaviour of band partitioning as a function of synchronous frequency we used the Fortran code *Transvrs* [12] driven with by a Mathematica input to the data set to calculate the kick factors and synchronous frequencies. The results of this calculation are shown in Fig 5 for $a/\lambda$ given by 0.229 (a), 0.19 (b) and 0.161 (c), in which the kick factors are calculated for structures with a phase advance ranging from 120 to 180 degrees. The general trend for the first three bands is quite clear, namely, rather independently of the iris dimension, the second and third dipole bands are enhanced at the expense of the first band as the phase advance per cell increases from the initial value of $2\pi/3$. The effect of finite group velocity on the kick factor has been left to a later publication as until recently the code was unable to incorporate this effect. However, inclusion of the finite group velocity does not modify our general conclusions on the partitioning of modes.

In conclusion, the present NLC design limits $a/\lambda \sim 0.18$ and thus the first three bands of the SW structure will be required to be damped and detuned. For the $5\pi/6$ structure only the first dipole bands must be damped and the cell frequencies detuned together with the third band which will be required to be detuned and moderately damped or not damped at all. Further studies are in progress on assessing the damping requirements of the $3^{rd}$ dipole band in the $5\pi/6$ structure.

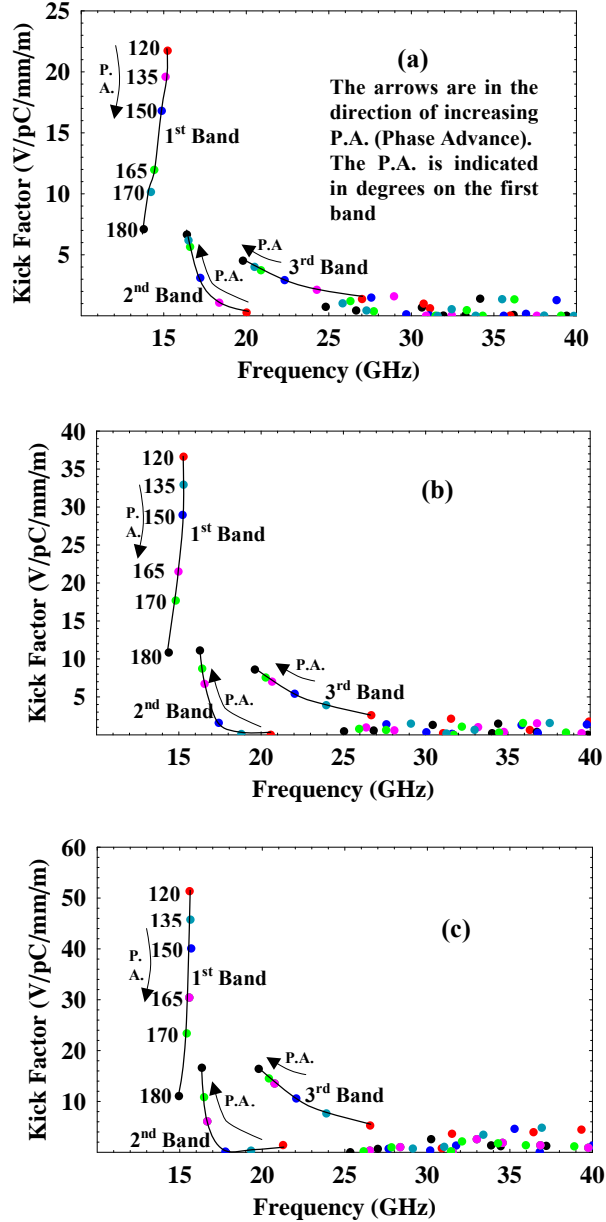

Figure 5. $K_n$ as a function of synchronous frequency for several irises radi: (a) 6mm, (b) 5 and (c) 4.23 mm.